\documentclass[prl,showpacs,twocolumn,preprintnumbers]{revtex4}
\usepackage{graphicx}
\usepackage{dcolumn}
\usepackage{amsmath}
\usepackage{bm}

\begin{document}

\title{The ``Quantum Mousetrap'':  \\ Entangled States and Gravitational Waves.}

\author{Fabrizio Tamburini$^\dag$}

\author{Bruce A. Bassett $^\ddag$} 
 
\author{Carlo Ungarelli$^\dagger$}

\affiliation{$^\dag$ Dept. of Astronomy University of Padova, Vicolo
dell'Osservatorio 3, IT-35122 Padova, Italy.}

\affiliation{$^\ddag$ SAAO / UCT, Observatory Road
Observatory, Cape Town 7925 South Africa}

\affiliation{$^\dagger$CNR IGG Pisa, Via G. Moruzzi 1, 56124 - Pisa, Italy}

\begin{abstract}
We propose a ``\textit{thought technique}'' for detecting Gravitational Waves (GWs)
using Einstein--Podolski--Rosen (EPR) photon Entangled States. 
GWs decohere the entangled photon pairs, introduce a relative rotation 
and de--synchronize Alice and Bob's reference frames
thus reducing the measured non-locality of correlated quanta 
described by Bell's inequalities.
Gravitational Waves, distorting quantum encryption
key statistics away from a pure white noise, act then as shadow
eavesdroppers. The deviation from the intrinsic white-noise
randomness of a Quantum Key Distribution process can reveal the
presence of a gravitational wave by analyzing the emerging color
distortions in the key. Photon entangled states provide the key
advantage of revealing the polarization rotation introduced by GWs 
without the need of previously fixed reference frames.
\end{abstract}

\pacs{03.65.Ud, 04.80.Nn, 03.67.-a}
 
\maketitle

\section{Introduction}
Quantum cryptography provides a stunning application of
Einstein-Podolsky-Rosen (EPR) correlations and Bell's inequalities
\cite{bb84,BB92,ekert91,benn,bell,bou,jenn99}. 
Not only does it promise perfectly
secure key distribution but we argue that it may also  allow the
detection of the shadowy traces of gravitational waves whose
existence is one of the most important outstanding tests
of Einstein's General Relativity and the subject of a number of current
and next-generation experiments \cite{trap1,mensky2, weber,LIGO,VIRGO,GEO,LISA,maggiore,cruise,montanari}.
Even if until now a clear evidence of GWs has not been obtained yet 
with nowadays detectors, an indirect proof of their existence was given in
the 1970's by the radio-astronomical observations of Hulse-Taylor
pulsar \cite{hulsetaylor}.

A general Quantum Key Distribution (QKD) scheme
consists of a random key generation by two parties A and B (Alice and Bob)
interested in communicating securely.
By performing a sequence of measurements on these entangled pairs of photons,
decided by a previously chosen protocol (BB84, B92)  \cite{bb84,ekert91,BB92},
Alice and Bob  determine the key they will use to encrypt their
message. 
An attack may be made by an eavesdropper (Eve) who secretly attempts 
to determine  the key as the pairs of entangled EPR quanta travel 
to Alice and Bob. 
The vital advantage that quantum mechanics
provides lies in the \emph{impossibility} that an eavesdropper
(Eve) can intercept the secret key without giving away her
presence to Alice and Bob, since such interception unavoidably
alters  the entanglement of EPR pairs.

Variants of the standard BB84 protocol
based on the transmission of single pairs of EPR photons have been used
recently in practical quantum key distribution over optic fiber networks
more than a hundred of km in length \cite{hiskett06}.  
Similar experiments have  illustrated the feasibility of quantum
encryption and single photon exchange in practical situations 
outside of a research laboratory and in space--to--ground links
\cite{jenn99,aspel,aspel2,pavil1,pavil2}.

Lorentz transformations, quantum metric fluctuations,
gravitational waves and, more in general, gravitational fields, 
decohere and introduce nonunitarity in a single qubit state and
in entangled states\cite{ginrichadami03,bergou03, terashima03, peres04, trap1}
but a unitary, precise, theoretical description is still missing.
Now the field is sufficiently mature to be a tool in
fundamental research beyond the foundations of quantum mechanics.

In this paper we propose a simple \textit{gedankenexperiment}, a ``thought technique'', 
to detect the effects of gravitational waves through the distortions they cause
in the statistics of the quantum keys determined by Alice and Bob.
The interaction of the EPR pairs, shared between Alice and Bob, with the
ripples of spacetime would produce similar effects to Eve's attacks.

\section{Photon polarization states and GWs}

We start considering the following thought experiment, in which
Alice and Bob, in their reference frames $(x^0_A, x^1_A, x^2_A, x^3_A)$
and $(x^0_B, x^1_B, x^2_B, x^3_B)$, having null relative
velocities in a Minkowsky spacetime, 
share entangled photons in the singlet state of eqn. (\ref{eq1})
\begin{equation}
|\Psi\rangle =\frac{1}{\sqrt{2}}[|H\rangle _{A}|V\rangle _{B}-|V\rangle
_{A}|H\rangle _{B}].
\label{eq1} 
\end{equation}
The subscripts $| ~\rangle _{A}$ and $| ~\rangle _{B}$ indicate that the 
physical quantities are measured in the reference frame of Alice or Bob's respectively.
The two parties are separated by the distance $l_{AB}$, 
measured along the common direction of $x^{1}_A$ and $x^{1}_B$ axes.
that has the same length in both reference frames. 
The axes $x^{2}_{A,B}$ and $x^{3}_{A,B}$ coincide with the vertical 
and horizontal directions $|V\rangle _{A,B}$  $|H\rangle _{A,B}$ 
of photons' polarization directions for both reference frames.

Let's consider for the sake of simplicity, and, without loosing in generality,
the ideal case in which a monochromatic plane GW, 
 \begin{equation}
h^{(TT)}_{\mu \nu}=\left(
\begin{array}{cccc}
0 & 0 & 0 & 0 \\
0 & h_{+}\cos \Theta  & h_{\times }\cos \Theta  & 0 \\
0 & h_{\times }\cos \Theta  & -h_{+}\cos \Theta  & 0 \\
0 & 0 & 0 & 0
\end{array}
\right)
\end{equation}
is propagating along the negative $z$ direction in Bob's reference frame.
The wave equation is expressed in the Transverse-Traceless (TT) gauge.
$\lambda_{GW} \gg l_{AB}$ is the wavelength, $h_{\times }$ and $h_{+}$
are the amplitudes of the GW's polarization modes ``$\times$'' and ``$+$'', 
$\Theta =k_{\mu}\cdot x^{\mu }_B$ is the phase of the wave at the position 
$x^{\mu }_B$ and $k^{\mu }=(\frac{2\pi }{\lambda _{g}},0,0,-\frac{2\pi
}{\lambda _{g}})$ the GW's wave-vector. 
More in general, while Alice remains in a Minkowsky spacetime having galilean metric tensor 
$g^{(0)}_{A, \mu \nu}$, Bob's region is described by a perturbed metric
$g_{B, \mu \nu}=g^{(0)}_{\mu \nu}+h_{\mu \nu}$.

Alice and Bob then perform Bell state measurements with randomly swapping 
polarizers. If a polarizer happens to be  correctly oriented, the incident photon
is detected and a ``1'' is recorded, otherwise a ``0''. Repetition generates two
equal length  binary strings  $K_A$ and $K_B$,
corresponding to the measurements of Alice's and Bob's detectors.

Alice and Bob then publicly announce the orientations of
their polarizers corresponding to each element in $K_A$ and $K_B$.
They then eliminate the elements  of $K_A$ and $K_B$ corresponding
to non-coincident orientations of the two polarizers. The string entries
of the remaining subsets of $K_A$ and $K_B$  form the
two quantum keys, $k_{A}$ and $k_{B}$. In
the absence of gravitational waves and noise the two keys coincide,
$k_A = k_B$,  since  the  photon pairs are perfectly  entangled.
After having built the key $k_A$, Alice proceeds with the transmission of an 
encrypted message to  Bob, who decodes it with his key, $k_B$.
However, this is not of interest to us now. Instead, cross-correlation
of the keys $k_A$ and $k_B$ allows, in principle, the detection of
gravitational waves.

This detection proceeds thanks to
a fundamental property of quantum cryptography: the key $K=k_A \otimes k_B$ 
of an ideal experiment is a Markovian process with zero--memory 
step, written in terms of a pure white noise--generated discrete random sequence of 
``0''s and ``1''s \cite{gisin2002}. 
The presence of a gravitational wave colours the cross-correlation statistics 
so the the keys are no longer ``white''.
So, for the strings $k_{A}$ and $k_{B}$, the probability of having a ``1'' (a
detection) is no longer equal to the probability of a ``0'' (a
non-detection). In addition the two strings will no longer coincide element 
by element: $k_A \neq k_B$. 
 
Seen from a classical point of view, a gravitational wave introduces a 
discoloration in the quantum key by changing the 
arrival time of the photons at Alice and Bob, by altering the detectors'
local time and the path length, $ l_{AB}$, travelled by the photons. 
Another relevant macroscopic effect is observed in Bob's $x^2_B$ and $x^3_B$ 
axes, that are misaligned with respect  to the correspondent ones 
of Alice's reference frame because of a rotation $\Delta \theta \propto h$.
At the quantum level, instead, the interaction of the GW with the photons 
of the entangled pairs can be described in terms of the interaction between 
a bath of coherent  states of gravitons $|g\rangle$ \cite{kok03} and photons, 
via elementary graviton--photon scattering processes (e.g. elastic 
gravitational Compton scattering $g\gamma \rightarrow g\gamma $)
that are shown to cause decoherence, but at the second order in 
the GW amplitude $h$ \cite{scadron,cho95,dlogi}.

Being always $h^{2}\ll h$, we focus on the effects that can cause a deviation on the 
quantum key distribution at first order $h$.
The change of relative local times and the relative reference frame rotation of Alice 
and Bob's will introduce a perturbation at the first order in $h$. 
To reveal the change of local times, the quantum cryptographic setup
would require a sensitivity better than the time coherence 
of the entangled photons and an extremely precise synchronization 
of the two reference frames. 

Generally, the detection coincidences dramatically depend on the synchronization 
of Alice and Bob's reference frames and it is still not clear whether entanglement 
could always be an intrinsic invariant relativistic property of the pair \cite{jor06,paw07}.
Crucial effects such as geodesic deviations of entangled quanta in a curved spacetime, 
the problem arising from synchronization of the two observers and from the 
coupling terms and of the potentials existing between the entangled quanta
(e.g. the effect of a Coulomb potential between two entangled ions), 
have not been yet completely faced in a fully relativistic approach. 
For this reason the detection of gravitational waves with entangled atom 
interferometers realized with massive quanta must take in account all of these 
important effects because 
the ions (or atoms) experience the presence of a non--negligible inter--particle
potential that do not have still a clear relativistic description \cite{alba07}.
Entangled photons may provide a simpler approach to this problem:
quantum field effects such as photon--photon interaction and the interaction 
potential between two entangled photons traveling along opposite directions 
can be neglected in our approach, since they give terms $\propto h^2$ or even
at higher order.

Neglecting at this first stage also the effects due to the lack of time synchronization
between Alice and Bob  (see e.g. \cite{trap1}), we focus now
on the disentanglement due to the rotation $\Delta \theta$ of the two reference
frames. The parallel transport of the polarization vectors' direction between
A and B will then reveal the presence of a Gravitational Wave.
In order to analyze this effect one may construct the cross-correlation 
matrix between the two keys $k_{A}$ and $k_{B}$  and search for off-diagonal 
power (see Fig. 1). 

\begin{figure}
\includegraphics[width=8cm]{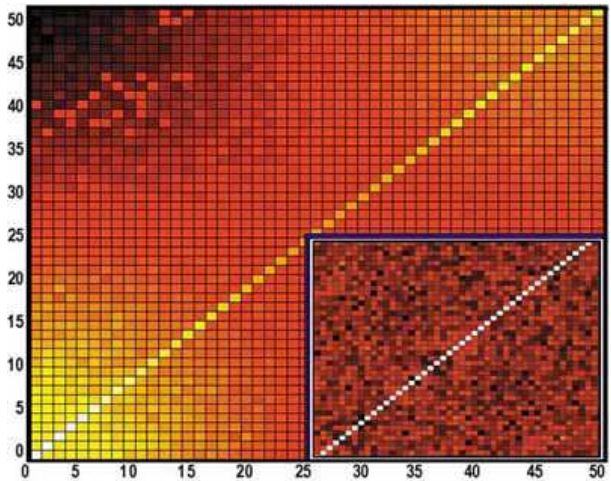}
\caption{The averaged cross-correlation matrix of sample
$50$-element keys $K_{A}$ and $K_{B}$. \textbf{Inset:} The
idealized white-noise case (without gravitational waves). The
diagonal dominates in the large key length limit where the
cross-correlation is simply $\propto \delta _{ij}$. \textbf{The
main figure} schematically shows the off-diagonal power
induced by a deterministic gravitational wave that
colours the white noise of the ideal uncorrelated and unperturbed
string.}
\end{figure}

By using the quantum key, we then define the {\it accumulated fluctuation} 
$\xi(t)$ as the absolute value of the difference, for a given temporal length $t$,
of the number of non-detections, $N_{[0]}$, and detections,
$N_{[1]}$, in $K$, {\em viz.} $\xi(t) \equiv |aN_{[1]}-bN_{[0]}|=0$,
where $a$ and $b$ are real numbers that depend on the angles used in the protocol.
This procedure is equivalent to the accumulation in time of the effects of GWs, 
making, in principle, their detection possible. 

Let's consider the general effect of the gravitational field
in more detail. Each polarizer is described by the states
$|H\rangle \cos \varphi_{A,B}+|V\rangle \sin \varphi_{A,B}$ where
the angles $\varphi _{A}$ and $\varphi _{B}$ swap between chosen
values. The probability that the Bell state overlaps with the
vector describing the polarizer is $P^{\Psi _{1}}=\frac{1}{2}\sin
^{2}(\varphi _{A}+\varphi _{B})$. If $\varphi _{A}=-\varphi _{B}$
and the two photons are detected the corresponding bit in the
string is set to ``1''; when $\varphi _{A}\neq \varphi _{B}$ or the
two photons are not detected a ``0'' is recorded. The probability of
having a ``1'' in the string is $P_{1}=\frac{1}{2}\sin ^{2}(\psi )$,
where $ \psi =2\varphi _{A}$ while the probability of having a ``0''
in the string is $ P_{0}=1-\frac{1}{2}\sin ^{2}(\psi )$. 
The standard angles used in the most common protocols
like BB84 and/or B92 (i.e. $0$, $\pm \frac{\pi}{4}$, $\pi$) 
show a dependence from the GW effect only at the second order in $h$.
This would make GWs far from being detectable with this
method.
By using instead the angles $\varphi _{A,B}=(-\pi /6,\pi /6)$, the detection and
non-detection probabilities are $P_{1}=\frac{3}{8}$ and
$P_{0}=\frac{5}{8}$, respectively, with a linear dependence from the GW's 
amplitude $h$.

In the absence of gravitational waves,
we build with our protocol a discrete zero-mean Markovian process 
$\xi _{A,B}(t)\equiv 5N_{[1]}(t)-3N_{[0]}(t)$.
By approximating the data record of length $T$ with a continuous
process, in the large number limit we obtain
\begin{equation}
\xi _{A,B}(T) \simeq \frac{\Gamma_{ph}}{2}\int_{0}^{T}dt (5P_{1} -3P_{0}),
\end{equation}
where $\Gamma _{ph}$ is
the detection rate of coincidences between entangled photon pairs. 
The presence of GWs induce a discoloration in $\xi _{A,B}(t)$ and in the
quantum key affecting the statistical properties of the
accumulated fluctuations. 

The detection and non-detection probabilities ($P_{1},P_{0}$) in the presence of a
gravitational wave become at linear order in the GW amplitude,
$P_{1}\simeq \frac{3}{8} + \frac{\sqrt{3}}{4}\Delta \theta$ and
$P_{0}\simeq \frac{5}{8}-\frac{\sqrt{3}}{4}\Delta \theta$,
being $\Delta \theta \propto h$. 
The accumulated fluctuation at time $T$ in the
long wavelength limit will become $\xi
(T)=2\sqrt{3} \Gamma_{ph}\int_{0}^{T}h(t)dt$. 
By assuming for example $\Gamma_{ph}\sim 10^5$, $h=10^{-18}$ and polarization
sensitivity $10^{-10}$ \cite{pvlas}, we would obtain a signal of $0.0035$ qubits/sec 
emerging from a pure white noise data record (i.e. $\sim 12$ qubits after 
$1$ hour of integration). It is easy to infer that, even if the probability
of detecting a GW is not null, to obtain a realistic estimate of GWs' effects
we would need an extremely low--noise setup.

In the idealized case where complex and experiment-specific noise
sources (such as thermal and seismic fluctuations influencing also 
the fasten axis of the polarizers) are neglected,
the intrinsic fluctuations in the time series due to the
fluctuations of the polarization directions
are described by a frequency-independent random process
characterized by the noise spectral density  - the
noise-induced mean square fluctuations per unit frequency that can
be modeled for a precise setup and then measured after the construction. 

In the following, we estimate the noise level due to the uncertainty in 
the measurement of the polarization angle of the entangled photons. For the 
sake of simplicity we will assume that each element of the qubit series sent 
to both Alice and Bob is affected by white noise, represented by a gaussian 
process with zero mean and variance equal to the error $\Delta\theta$ 
in the polarization measure: 
$\langle n_I \rangle =0\,$ and
$\langle n_I n_J\rangle =\delta_{IJ}(\Delta\theta)^2 \,$
(I,\,J=1\dots N),
where $N=N_{c}/2$ and $N_c$ is the number of couples of entangled photons 
so far detected.

In order to compare the noise floor to a typical value for current 
earth-based detectors we now compute the noise spectral density (i.e. the 
noise power per unit frequency). Taking the discrete Fourier transform of 
the noise  
\begin{equation}
\tilde{n}_J=\delta t \sum_{I=0}^{N-1}e^{-\frac{2\pi I J}{N-1}}n_I
\end{equation}
where $\delta t$ is the sampling time of the qubit series, it is straightforward 
to show that 
\begin{equation}
\langle \tilde{n}_I\tilde{n}^*_J\rangle = 
\left[\delta t (N-1)\delta_{IJ}\right]\delta t (\Delta\theta)^2.
\end{equation}
The quantity between square brackets it is the discretization of the 
Dirac delta function in Fourier space so the noise spectral density for 
each element of the qubit series is $\tilde{S}_n=\delta t (\Delta \theta)^2$.
We now divide the qubit series into $M$ segments each containing $L$ qubits such that 
$(M \cdot L=N=N_c/2)$.
For each segment the noise spectral density is reduced by 
a factor $L$ with respect to the value corresponding to a single qubit so 
\begin{equation}
\tilde{S}^{(J)}_n=\frac{1}{L}\delta t (\Delta \theta)^2, \,\,\,\, (J=1, \dots, M).
\end{equation}
Assuming that the production rate $\Gamma_{\rm ph}$ 
of entagled pairs of photons is equal to the detection rate 
($\Gamma_{\rm ph}=2\delta t^{-1}$), we obtain 
\begin{equation}
\tilde{S}^{(J)}_n=\frac{4}{\tau}
\left(\frac{\Delta \theta}{\Gamma_{\rm ph}}\right)^2, \,\,\,\, (J=1,\dots, M),
\end{equation}
where $\tau$ is the time duration of the segment.  
So the spectral amplitude of the noise 
$\tilde{h}_n=\sqrt{\tilde{S}_n}$ for each segment is 
\begin{multline}
\tilde{h}^{(J)}_n=2\times 10^{-23}
\left(\frac{\Delta\theta}{10^{-10}{\rm rad}}\right)
\\
\left(\frac{10^6\,{\rm s}}{\tau}\right)^{\frac12}
\left(\frac{10^{10}\,{\rm s}^{-1}}{\Gamma_{\rm ph}}\right)\, {\rm Hz}^{-\frac{1}{2}}
\end{multline}
to be compared with the best LIGO design sensitivity (around 100Hz) \cite{ligo2}, 
$\tilde{h}_n\sim 3\times 10^{-23}{\rm Hz}^{-1/2}$. 

Since those quantum entangled states based detectors use the
properties of photon-photon correlations of each single entangled
pair, the accumulated fluctuation is a random process with zero
mean and linearly increasing variance.
However this detection method can
be better improved by applying the technique of the randomness of the
choices of measurement basis by Alice and Bob \cite{basis}, which
gives significant advantages in cases where the qubit error rate is crucial.
What we have shown is that gravitational waves will act as \textit{shadow
eavesdroppers}, reducing the degree of entanglement between quantum
states controlled by Bell's inequalities.

\section{Discussion and Conclusions}
We have outlined how quantum technology may be exploited to yield
a potentially sensitive detector of gravitational waves. The setup (``Mousetrap'')
can measure angular deviations that are first order in the GW
amplitude $h$ and, we point out that, counting the coincidences between 
detections of the EPR pairs the setup will not be affected by laser's shot noise. 
This makes our setup a good candidate to detect high frequency GWs, 
typical of string--cosmology scenarios \cite{mca08}.

The advantages of the Mousetrap with respect to a classical GW detector based on
polarization change of an electromagnetic wave \cite{cruise} is provided 
by the properties of twin-- and entangled--photon ellipsometry that are part
of a QKD scheme. This permits to estimate with high precision
the deviation $\Delta \theta$ of the polarization vectors in each pair accumulated 
during the travels from the source to the observers Alice and Bob.
This procedure may improve the detection of GWs especially in situations
when the orientation between Alice and Bob's reference frames 
becomes crucial, like in space--based experiments.
With quantum ellipsometry techniques, in fact, the experimenter does not need 
to fix the direction of its reference frame (e.g. with respect to the stars)
and ideally does not even need to characterize for each measurement 
its optical setup, reaching in principle the quantum limit
\cite{sergienko, sergienko02, kok03}.
The crucial additional advantage of using entangled states instead of
twin (or correlated) photons is that the building of the cryptographic
keys permits the use of stochastic techniques to accumulate the
effects of GWs and, also, to detect the presence of a stochastic
gravitational wave background.
This setup could benefit or give advantages to atom interferometry
techniques recently proposed \cite{dim08,tin07}.

The use of $n$ GHZ states may help to to beat the shot noise limitation,
with its $1/n$ dependence, but with the disadvantage of having a fainter 
and fainter source the more complex is the GHZ state \cite{bagan,zeibook}. 
Finally, entanglement and teleportation may be used to propose other more
sophisticated GW detectors, as those based on the communication of spin, 
of directions and of whole reference frames in different points of spacetime 
\cite{massarpopescu95,gisinpopescu99, bagan00, bagan, bagan2,peresscudo01a,
peresscudo01b}, but all these pioneering techniques still require a too critical and difficult
local control of the quantum states, a fine tuning of the coupling between 
the quanta and the communication of different quantum states over large 
distances that would unavoidably strongly affect their performances.

In principle our \textit {gedankenexperiment} could represent an alternative
approach to the already existing interferometric detectors in the high
frequency domain. 
Our estimations show that the noise level of this ideal setup is of
the same order of magnitude as in the best LIGO design sensitivity.
Anyway a real, detailed, understanding of all relevant noise sources 
is still lacking and depends on exact details of the detector set-up 
and on the future development of detector and source technology. 
But this goes beyond the purpose of this paper.

\acknowledgments{
We thank Markus Aspelmeyer, Cesare Barbieri, Naresh Dadich, 
Thomas Jennewein, Luca Lusanna, Tony Rothman and Alberto 
Vecchio for helpful discussions and insightful comments on the manuscript.}

\end{document}